\documentclass{sig-alternate-per}
\usepackage[utf8]{inputenc}
\usepackage{amsmath, amsfonts}
\usepackage{enumitem}
\usepackage{dsfont}
\usepackage[hyphens]{url}
\usepackage{hyperref}
\usepackage{makecell}
\usepackage{multirow}
\usepackage{booktabs} 
\usepackage{graphicx}
\usepackage{subcaption}
\usepackage{comment}
\usepackage{xspace}
\usepackage{color}
\usepackage{flushend}
\usepackage{arydshln}
\usepackage{amsmath}
\usepackage{xcolor}
\usepackage{color, colortbl}

\usepackage[nocompress]{cite}
\usepackage{glossaries}
\makeglossaries

\newacronym{poi}{PoI}{Point of Interest}
\newacronym{lbsn}{LBSBN}{Location-Based Social Networking}
\newacronym{grm}{GRM}{Group Regularity Mobility Model}
\newacronym{bfs}{BFS}{Breadth-First Sampling}
\newacronym{dfs}{DFS}{Depth-First Sampling}
\newacronym{ppmi}{PPMI}{Positive Pointwise Mutual Information}
\newacronym{cdf}{CDF}{Cumulative Distribution Function}
\newacronym{cv}{CV}{Coefficient of Variation}

\begin{document}

\conferenceinfo{Workshop on AI in Networks and Distributed Systems (WAIN) 2021}{~~~Milan,Italy}
\CopyrightYear{2021}

% \title{Understanding network mobility: A node embedding based approach}
\title{Understanding mobility in networks: A node embedding approach}
% Characterizing the human mobility: A node embedding based approach
% Modeling and tracking the human mobility: A node embedding based approach
% Modeling and analyzing human mobility: A node embedding-based approach.

\numberofauthors{6}
\author{
%
% 1st. author
\alignauthor
Matheus F. C. Barros\\
      \affaddr{Universidade Federal de Ouro Preto}\\
     
% 2nd. author
\alignauthor
Carlos H. G. Ferreira\\
      \affaddr{Universidade Federal de Ouro Preto\\Universidade Federal de Minas Gerais\\Politecnico di Torino}\\
     
% 3rd. author
\alignauthor
Bruno Pereira dos Santos\\
      \affaddr{Universidade Federal de Ouro Preto}\\
     
% 4th. author
\and  
\alignauthor
Lourenço A. P. Júnior\\
      \affaddr{Instituto Tecnológico de Aeronáutica}\\
     
\alignauthor
Marco Mellia\\
      \affaddr{Politecnico di Torino}\\

\alignauthor
Jussara M. Almeida\\
      \affaddr{Universidade Federal de Minas Gerais}\\
}
     
%Conference

%

\maketitle

\begin{abstract}
Motivated by the growing number of mobile devices capable of connecting and exchanging messages, we propose a methodology aiming to model and analyze node mobility in networks. We note that many existing solutions in the literature rely on topological measurements calculated directly on the graph of node contacts, aiming to capture the notion of the node's importance in terms of connectivity and mobility patterns beneficial for prototyping, design, and deployment of mobile networks. However, each measure has its specificity and fails to generalize the node importance notions that ultimately change over time. Unlike previous approaches, our methodology is based on a node embedding method that models and unveils the nodes' importance in mobility and connectivity patterns while preserving their spatial and temporal characteristics. We focus on a case study based on a trace of group meetings. The results show that our methodology provides a rich representation for extracting different mobility and connectivity patterns, which can be helpful for various applications and services in mobile networks. 
\end{abstract}

%https://github.com/jbenet/latex/blob/master/acm/sig-alternate.tex

\section{Introduction}

Analyzing and modeling mobility patterns have fundamental roles in mobile network prototyping, design, and deployment. To study such aspect, we can use mobility data gathered directly from the devices or the infrastructure that they are attached to. Examples of such data are GPS trace, call detail records (from phones), location-based social media, etc~\cite{Hess:2016}. With these data, one is able to study the behavior of mobile entities and model their mobility patterns, which can be used to evaluate the performance of newly designed networking protocols and solutions~\cite{Ferreira:2016, Nunes:2017, santos2018mobile}.   
% Analyzing and modeling mobility patterns have fundamental roles in mobile network prototyping, design, and deployment. To study such aspect, we can use mobility data gathered directly from the devices or the infrastructure that they are attached to. Examples of such data are GPS trace, call detail records (from cell phones), location-based social media, etc~\cite{Hess:2016}. With these data, one is able to study the behavior of mobile entities by modeling their mobility patterns, which can be used to evaluate the performance of newly designed networking protocols~\cite{Nunes:2017} or build new solutions~\cite{santos2018mobile, reis2021mobetx}.   

One way to model mobility is through traditional temporal graph theory. In this approach, nodes are the network entities, and the edges are interactions (i.e., calls, message or data exchanges) over time. Conveniently,  well-known graph theory models~\cite{Vastardis:2013, barabasi2016network} are ready to use. Armed with this, several proposals design solutions for, e.g., networking routing by exploring node importance criteria such as centrality metrics, community structure, and social metrics~\cite{Sobin:2016, Wan:2017, Zhang:2017,  santos2020routing, Guo:2021, borgatti2006graph, barabasi2016network}.  %Indeed, several criteria of node connectivity importance can be found in the literature, such as Closeness, Eccentricity, Eigenvector, Betweenness Centrality, etc~\cite{borgatti2006graph, barabasi2016network}.
However, each metric captures a specific notion of node importance, indicating which nodes are central elements in the network structure. Although some earlier studies and solutions on network mobility have explored mobility modeling based on such metrics, they cannot generalize a network nodes' importance criteria and, most important, do not consider the temporal dimension~\cite{Valente:2008, Ferreira:2019}. These limitations are especially harmful, for example, to the validation of opportunistic routing protocols since none of the existing modelings approaches fully cover the nodes' importance criteria on the network connectivity~\cite{Nunes:2018,santos2018mobile}.
% Indeed, several criteria of node connectivity importance can be found in the literature, such as Closeness, Eccentricity, Eigenvector, Betweenness Centrality, etc~\cite{borgatti2006graph, barabasi2016network}. However, each such metric captures a specific notion of node connectivity importance that indicates nodes are most important or central in a network for solutions designs. Although some earlier studies and solutions on network mobility have explored the modeling mobility based on specific metrics, they cannot generalize a network nodes' importance criteria and, typically, do not consider the temporal dimension~\cite{Valente:2008, Ferreira:2019}. This limitation is especially harmful, for instance, to the validation of opportunistic protocols or routing protocols since none of the existing modelings approaches fully cover the nodes' importance criteria on the network connectivity, typically helpful in studying and building solutions on mobile networks~\cite{Nunes:2018,santos2018mobile}.

Recently, \textit{node embedding} techniques have been proposed as an alternative to studying mobility rather than traditional graph theory. There exists a range of mobility-related applications of embeddings such as \gls{poi} prediction and recommendation~\cite{yang2019revisiting, yang2020lbsn2vec++, rahmani2019category, Agrawal:2021}, and urban mobility modeling and planning \cite{zhang2021unveiling, Wang:2019, Xu:2020}. In a nutshell, \textit{node embedding} encodes a graph structure to low dimensional vector space (a.k.a. latent space), unveiling hidden patterns that are hard to catch using only the graph topology~\cite{Lu:2019}. Although such features are desirable for exploring mobility patterns to build solutions in a network context, few efforts use \textit{embeddings} to understand mobility patterns and capture the importance of connections established by mobile nodes, specifically valuable to design routing and information dissemination solutions in mobile networks~\cite{Ferreira:2019}.

Aiming to address the aforementioned issues, we propose using \textit{node embedding} to model and analyze the mobility pattern over spatial and temporal dimensions. %As prior efforts focused on topological only and specific properties that are not able to catch the overall notion of connectivity importance as nodes change their connections in the network.
Our \textit{embedding}-based methodology goes beyond capturing the topological aspects of the network, the mobility, and it factors the evolution of such patterns through a sequence of networks over time too. We present a case study related to the regular meeting of groups of people that applies our approach. The results show that we can extract and analyze mobility patterns, identifying strategical nodes for network connectivity and tracking the evolution of such properties over time. Our approach, albeit preliminary, can be seen as a first alternative to support the development of communication protocols, framework design, share and allocation of network resources, or applications~\cite{Vastardis:2013, Xia:2016} which considers node stability over time.

This work is organized as follows: Section~\ref{sec:related} summarizes the main efforts in the literature that employ \textit{embeddings} approaches for studying network mobility, Section~\ref{sec:dataset} presents our case study, while Section~\ref{sec:methodology} describe the methodology. Next, the results are presented in Section~\ref{sec:results}. Finally, Section~\ref{sec:conclusion} discusses the main findings, offers conclusions, and future directions.

\section{Related Work}~\label{sec:related}

% Modeling mobility and interactions between entities has rich applicability. In this sense, mobility-aware solutions were proposed focusing on routing metrics~\cite{Batabyal:2015, reis2021mobetx, santos2018mobile}, protocols and actions for coordination~\cite{Alajeely:2018}, modeling and analysis of the behavior of individuals on a mobile network~\cite{Hess:2015}. However, solutions supported by \textit{embeddings} techniques have been little explored yet. Our focus here is to model user mobility on mobile networks relying on concepts and models from network science, including network embedding as fundamental tools to drive our study. This section briefly reviews relevant prior work adopted approaches to model mobility patterns based on network embedding methods.

Modeling mobility and interactions between entities has rich applicability. For example, mobility-aware solutions were proposed focusing on routing metrics~\cite{Batabyal:2015, santos2018mobile}, protocols and actions for coordination~\cite{Alajeely:2018}, and analysis of individuals' behavior on a mobile network~\cite{Hess:2015}. However, solutions based on \textit{node embedding} techniques have been little explored. Following, we review some prior work that model mobility pattern using embedding methods. %This section briefly reviews relevant prior work adopted approaches to model mobility patterns based on network embedding methods.

Yu et al.~\cite{yu2018inferring} proposed inferring the relationship strength between users as well as understanding the motivation behind them. To do this, they model a contacts frequency network between two users whose edge weight is adjusted based on contextual information from the contact's location. Yang et al.~\cite{yang2019revisiting} focused on the tasks of predicting new friendships and predicting the location of users. First, the authors proposed LBSN2Vec, an embedding approach that receives as input a network of contacts modeled as a hypergraph from Location-Based Social Networking (\gls{lbsn}) data. In another work \cite{yang2020lbsn2vec++}, the authors introduced contextual information made possible by the creation of hyperedges considering not only users' contacts but also points of interest (\gls{poi}) and time.

Embeddings techniques in the context of mobility have been applied in information retrieval tasks. Specifically, in the context of \gls{poi} recommendation, Rahmani \textit{et al.}~\cite{rahmani2019category} proposed CATAPE, a method that simultaneously incorporates sequences of user locations and categories of \gls{poi}'s visited by them to recommend potential new points of interest. Likewise, Yuan \textit{et al.} combined information from historical and contrasting users' queries to find mobility similarity~\cite{yuan2020spatio}. Such approach focuses on recommending a \gls{poi} to a user through keywords in an incomplete query. Wang \textit{et al.} focused simultaneously on recommending and predicting users' \Glspl{poi}~\cite{Wang:2019}.% Their approach considers a sampling process in a heterogeneous graph representing time information, users, POIs, and their categories.

Motivated by urban mobility planning, Zhang \textit{et al.}\cite{Zhang:2019} used an embedding strategy to extract mobility patterns on transport lines using user mobility data in the transport system. Later, the authors considered data from \gls{poi}'s for the construction of urban mobility networks for users in order to find mobility patterns among users considering both data sources~\cite{zhang2021unveiling}. Similarly, Wang \textit{et al.} proposed a framework based on embedding whose focus is to analyze the dynamics of communities formed by mobility between \glspl{poi} in different periods (i.e., weekdays and weekends)~\cite{Wang:2018}.

% Já o trabalho de Xu \textit{et al.} \cite{Xu:2020} apresenta o \textit{SUME}, que é uma estratégia baseada em \textit{embeddings} para inferência demográfica de usuários a partir POIs. O \textit{SUME} recebe como entrada um modelo de rede heterogênea em que diversas categorias de arestas entre usuários e locais são criadas. Uma parte delas codifica co-visitações e a outra parte informações semânticas dos POIs. Em seguida, essas redes são representadas no espaço vetorial cujo objetivo é preservar a semelhança dos usuários nos padrões de mobilidade urbana. 

Unlike some previous work~\cite{yu2018inferring, yang2019revisiting, yang2020lbsn2vec++}, our approach does not consider contextual information. We notice that such information offers greater possibilities for pattern analysis and extraction. On the other hand, it poses extra challenges such as data collection, storage, and processing. In addition, collect such data demands financial, operational, privacy, and security issues. Therefore, unlike these works, our approach is independent of contextual information and is based exclusively on a sequence of contact graphs that represent the topological organization of the network. This is motivated by the problem addressed here, which consists of modeling the importance of users for network connectivity considering their mobility. Therefore, we only use network' topological information in this initial effort, although we can extend it to consider contextual information.

% Nós também observamos que algumas abordagens adotam estratégias de amostragem através de passeios aleatórios, similar a nossa proposta, apesar de considerarem grafos heterogêneos (múltiplas categorias de nós e/ou de arestas)~\cite{Wang:2018, Wang:2019, yang2020lbsn2vec++}. De fato, passeios aleatórios tem possibilitado identificar importantes propriedades acerca dos nós no grafo. Por exemplo, conectividade entre os nós (proximidade) e nós que desempenham papeis similares de conectividade (\textit{Hubs} e periféricos~)~\cite{Grover:2016}. Na abordagem aqui proposta, um grafo tradicional e homogêneo foi usado no processo de amostragem, isto é, nós representam usuários e arestas representam o contato entre estes usuários. Além de não utilizarmos informações contextuais, a dimensão temporal é considerada através de uma sequência de grafos modelando janelas de tempo (\textit{snapshots}) distintas, que nos permite avaliar a dimensão espaço-tempo. 

% In terms of applications, previous works focusing on prediction \cite{yang2019revisiting, yang2020lbsn2vec++} and \gls{poi} recommendation~\cite{rahmani2019category, Agrawal:2021, yang2020lbsn2vec++} are orthogonal to our effort as they are based on contextual information. 
Finally, we discuss works related to improving urban mobility and networking design~\cite{zhang2021unveiling, Wang:2019, Xu:2020}. Concerning these efforts, our proposal can be seen as an initial attempt towards planning and designing applications inherent to effective routing in opportunistic networks, which offers a range of applications in smart cities such as dissemination/collection of information, location, vehicular communication, among others~\cite{cai2021survey, syed2021iot}.
% In terms of applications, previous works focusing on prediction \cite{yang2019revisiting, yang2020lbsn2vec++} and \gls{poi} recommendation~\cite{rahmani2019category, Agrawal:2021, yang2020lbsn2vec++} are orthogonal to our effort as they are based on contextual information. Finally, we discuss works related to improving urban mobility~\cite{zhang2021unveiling, Wang:2019, Xu:2020}. Concerning these efforts, our proposal can be seen as an initial attempt towards planning and designing applications inherent to effective routing in opportunistic networks, which offers a range of applications in smart cities such as dissemination/collection of information, location, vehicular communication, among others~\cite{cai2021survey, syed2021iot}.
\section{Dataset}\label{sec:dataset}

One of the greatest challenges of performing mobility analysis in networks is the lack of reliable and open data in such field. This is due to privacy issues, scale (space, time, \# entities) or costs needed to collect and store such data (typically huge amounts). Fortunately, efforts have been done to build mobility models able to generate synthetic trajectories for mobile entities by using real data. Synthetic approaches offer a cheap and enabler alternative for quick and scalable performance evaluation of networking solutions in a near real environment with different scales of space, time, and number of entities \cite{Yi:2015}. Also, synthetic models enable reproducibility and fair comparison of different methods.
Given this, in this work we check the potential of node embeddings using synthetic mobility traces. In detail, we use the \gls{grm}~\cite{Nunes:2017}. \gls{grm} models group meeting dynamics and regularity in human mobility. It simulates the human mobility pattern by considering the cyclical and regular behavior commonly exhibited by humans, superposing  the effect of sporadic groups meetings such as sport events, presence of places for leisure, parties, etc. The results is a trace where nodes moves, replicating the periodic behavior with the superposition of those occasional meetings. These features differentiate \gls{grm} from other models in the literature. 

To produce a trace (nodes trajectories), \gls{grm} receives as input the area, duration, number of nodes, and groups. Some statistical parameters offer the ability to generate different scenarios: (i) \textit{Group Meeting Time} that defines the time interval between meetings; such interval is governed by a power-law with exponential cut-off with parameters $\alpha_{gmt}$ and $\beta_{gmt}$; (ii) \textit{Group Meetings Duration} that defines the time for which nodes will spend together; this duration follow a truncated power law distribution of parameters $\alpha_{dur}$ and $\beta_{dur}$; (iii) \textit{Group Structure} that decides which nodes will be at each meeting; the grouping approach follows also a power law with exponential cuts tuned by $\alpha_{size}$ and $\beta_{size}$; (iv) \textit{Social Context} is a network of social edges (``friends'') between nodes; \gls{grm} consider social relationship into account because people have higher probability to attend meetings in which friends' nodes participate into. 
For this, \gls{grm} assumes that each group has a regularity factor defined by $K$. For example, groups with a factor $k=24h$ meet every 24, 48, or 72h. This parameter works as a multiplier that generates the periodic behavior of the traits.
Here, we use the same parameters as the \gls{grm}'s authors to generate our  dataset trace (see Table~\ref{tab:ParametrosEntrada}). It is worthy to highlight that the choice of this trace, in particular, does not prevent our approach from being used with other traces.

\begin{table}[t]
\scriptsize
\centering
\caption{\gls{grm} input parameters.}
\label{tab:ParametrosEntrada}
\resizebox{0.95\linewidth}{!}{%
\begin{tabular}{@{}lc@{}}
\toprule
\rowcolor[HTML]{FFFFFF} 
\multicolumn{1}{c}{\cellcolor[HTML]{FFFFFF}{\color[HTML]{333333} \textbf{Parameters}}} & \textbf{Values}                    \\ \midrule
\# of Nodes                                                                            & 100                                \\
\# of Groups                                                                           & 500                                \\
\begin{tabular}[c]{@{}c@{}}Sim. Duration\end{tabular}                                  & 87 days                            \\
K                                                                                      & 70\% -24h; 15\% -7 days; 15\% - 6h \\
Grid                                                                                   & 30 x 30                            \\
Cell Size                                                                              & 50m$^2$                            \\
$\alpha_{gmt}$                                                                         & 3                                  \\
$\beta_{gmt}$                                                                          & 30 days                            \\
$\alpha_{dur}$                                                                           & 3                                  \\
$\beta_{dur}$                                                                          & 30 days                            \\
$\alpha_{size}$                                                                        & 2.24                               \\
$\beta_{size}$                                                                         & 30                                 \\
Social Network                                                                         & Gaussian Random Partition~\cite{Brandes:2003}          \\ \bottomrule
\end{tabular}%
}
\vspace{-0.3cm}
\end{table}

\section{Methodology}\label{sec:methodology}

This section describes the methodology we propose in our study, starting with our network model (Section \ref{sec:Network_model}) followed by the spatial-time aware node embedding approach applied to model a latent space (Section \ref{sec:DynamicNode2Vec}).

\subsection{Network Model}\label{sec:Network_model}

To model the nodes' contact network observed in graphs, we first discretize the period analyzed in non-overlapping time windows $\{\Delta_t\}$ of duration 24\,h\footnote{The choice of daily time windows is based on the meetings periodicity presented in Table \ref{tab:ParametrosEntrada}, but can be configured according to the scenario. It should be noticed that 85\% of meeting occurs on a daily scale.} represented by the set $T=\{1, 2, ..., 87\}$. For each time window $\Delta_t$ we build an undirected graph $G_{\Delta_t}$, with $G_{\Delta_t}=(V_{\Delta_t}, A_{\Delta_t})$, such that $V_{\Delta_t}$ = \{$v_1, v_2, v_3,..., v_i$\} is the set of nodes representing individuals at the instant of time $\Delta_t$; and $A_{\Delta_t}$ = \{$a_{ 1}, a_2, a_3, ..., a_n$\} a set of edges connecting any two nodes $v_i$ and $v_j$ ($v_i \neq v_j$). We assume two nodes are connected if they are within a radius of 100 meters during the considered $\Delta_t$. In practice, this could represent an exchange or a potential exchange of messages between any two nodes for specific scenarios and applications such as those provided by the Wi-Fi 6~\cite{yaakop2017bluetooth,yin2019survey} technology. It is important to highlight that such parameters can be configured according to the application.

With the definition of the graph, we consider some purely structural measures typically in the literature in mobile network applications that provide knowledge of node importance connectivity. Here, we can compare them to the patterns captured from our methodology. Among the various possibilities, we selected some of the primary measures of centrality (Degree, Betweenness, Closeness, Eigenvector), and \textit{clustering coefficient}, previously adopted in various solutions in network management \cite{Sobin:2016, Wan:2017, Zhang:2017, Nunes:2018,  santos2020routing, Guo:2021}. For brevity, we recommend reading \cite{newman2005measure, barabasi2016network} for more details about such metrics.

\subsection{DynamicNode2Vec}\label{sec:DynamicNode2Vec}

We use our prior DynamicNode2Vec technique to extract mobility patterns from mobile nodes~\cite{Ferreira:2019}. Such technique is able to temporally represent nodes from a sequence of graph networks (as modeled in Section~\ref{sec:Network_model}) in a low-dimensional vector space (embeddings) while preserving the existing network properties (e.g. neighborhood).
% To extract mobility patterns from mobile notes in a sequence of networks as defined in the previous section, we used the technique proposed in \cite{Ferreira:2019} hereafter referred to as \textit{DynamicNode2Vec}. This technique allows representing the nodes of a given sequence of graphs in a low-dimensional vector space (embeddings) while preserving the existing properties of the graph.
\textit{DynamicNode2vec} uses a \textit{biased random walk} algorithm based on a state-of-art method for static \textit{node embedding} known as \textit{Node2Vec} \cite{Grover:2016}. The intuition behind those walks is to traverse the graph randomly and produce a sequence (i.e., vectors) of visited nodes. The extraction of several random sequences generates a sampling process, where the appearance of nodes depends on the original graph. The algorithm defines two parameters: (i) $nw$ defined as the number of walks (or extracted sequences) per node; (ii) $wl$ the number of steps of each walk (or the length of each sequence). 
% Firstly, \textit{DynamicNode2vec} uses a \textit{biased random walk} algorithm of a state-of-art method for static \textit{node embedding} known as \textit{Node2Vec} \cite{Grover:2016}. The main idea of this algorithm is to cross paths in the graph, returning a sequence of visited nodes from a sampling process performed from all nodes. Thus, the first two parameters to be defined comprise the number of walks per node ($nw$) and the number of steps each takes ($wl$). It is also possible to configure the type of sampling to be performed, which can follow a \gls{bfs} or \gls{dfs}.

To configure how the sampling will be performed, the algorithm can use \gls{dfs} or \gls{bfs} through the $p$ and $q$ parameters. In details, $p$ determines the probability of immediately returning to an already visited node; $q$ controls whether the tour is close to the origin node (for \gls{bfs}) or it goes further away (for \gls{dfs}).
Here, we are interested in \gls{bfs} technique, which captures the proximity of nodes using a neighborhood-first strategy. In a nutshell, by sampling preferably neighbor nodes, the generated walks tend to include all nodes that belong to the same portion of the graph. Therefore, we tend to encode the importance of connections between nodes once mapped in the low-dimensional vector space. For this, we set $p=1$ and $q=0.5$ to get this behavior (see authors recommendations in~\cite{Grover:2016}. 
% Here, we are interested in \gls{bfs} technique, which captures the proximity (neighborhood-first) and, therefore, the importance of connections between nodes to be encoded in low vector space. To achieve this goal, we need to set the parameters $p=1$ and $q=0.5$ as the authors recommend in~\cite{Grover:2016}. The $p$ parameter determines the probability of immediately returning to an already visited node the $q$ parameter controls whether the tour is close to the origin node (using \gls{bfs}) or it goes further away (\gls{dfs}).

% In the former, the neighborhood visited from a given origin node is privileged to the nearest nodes, while the DFS consists of sequentially sampled nodes at increasing distances. Here, we are interested in BFS-based sampling, which captures the proximity and importance of connections between nodes to be represented in vector space. To control this behavior, two other parameters are used, $p$ and $q$. The first determines the probability of immediately returning to an already visited node. In contrast, the second controls whether the tour is close to the origin node, exploring the same neighborhood (i.e., corresponding to BFS), or whether it should go further away, exploring other nodes (i.e., corresponding to DFS). Following the authors' recommendations for our goal \cite{Grover:2016}, we define $p=1$ and $q=0.5$.

Given we have a different graph at each time window $\Delta_t$, we generate the paths (walks) for each of these graphs. As result, nodes that are closer to each other in a given time window have a high number of shared neighbors, and important edges for the network connectivity are more likely to appear close together in the multiple paths sampled in the respective graph $G_{\Delta_t}$. After, \textit{DynamicNode2Vec} incorporates an optimization model that efficiently observes the node association and their changes over time as they appear on the sampled paths, mapping them into a latent space temporally equivalent \cite{yao2018dynamic}. In summary, this is achieved as follows. 

First, for each observed time window we compute a node co-occurrence matrix through a sliding window (of size five as suggested in \cite{yao2018dynamic}) that runs over the sampled paths. That way, it counts how many times two nodes are co-visited within that sliding window. The greater the co-occurrence between these two nodes, the closer they are in the network, and the more important is the connection between them. This matrix is known as \gls{ppmi}. At the next step, we create a compact representation in $d$ dimensions from each \gls{ppmi} matrix derived from the sequences of observed networks. This step is done by a low-rank matrix factorization obtained by solving an optimization problem. We consider two factors in the objective function design to handle, respectively, \textit{overfitting} ($\lambda$) and \textit{alignment} ($\tau$). Their definition is based on a training process. For the sake of brevity, we recommend reading the original reference for more detail \cite{Ferreira:2019}. For our study, we set the parameters as follows: $nw=4$, $nw=8$, $\lambda=50$, $\ tau=15$ and $d=50$.

With the obtained embeddings, we can use two metrics to quantify the degree of mobility of nodes on the network, and the importance of the connections established by them: \cite{Ferreira:2019, yao2018dynamic}:
\\
\\
\noindent \textbf{Cosine Distance}: Quantifies the similarity between two vectors in the same vector space. Given two vectors generated by a node (user) $v_i$ in any two time windows $\Delta{t_1}$ and $\Delta{t_2}$, the cosine distance is defined as $cos(x,y)= 1- x \cdot y /( ||x|| ||y||)$. The cosine distance ranges from 0 to 1. Values close to 0 indicate that the node is always close to the same nodes in the two compared time windows; values close to 1 indicate that the node has drastically changed its network of contacts in the period, with total different neighbors.
\\
\\
\noindent \textbf{Vector norm}: By definition, the more a node appears in the sampled paths in the graph, the greater is the vector norm of its vector, taking into account all vector space of observed time. This way, it is possible to identify those nodes that exhibit an important connectivity pattern since they are frequently inserted in the sampled paths. %The norm of a vector corresponding to a node $v_i$ at time instant $\Delta_t$ is defined by $\|\mathbf{u}_{{v_i}_{\Delta{t}}}\|$.
The main advantage of using the vector norm regarding the frequency with which nodes appear on the sampled paths is that it is more robust to sporadic disturbances \cite{yao2018dynamic}.
\\

\section{Results}\label{sec:results}

In this section, we present the main results of our work. First, we illustrates our  embedding representation  of the mobile network considering  space and time characteristics. Next, the node mobility and importance is analyzed through the case study are presented.
% In this section, we present the main results of our work. Section~\ref{sec:illustrating} illustrates how our approach represents such mobile networks considering user mobility from a space and time characteristics. Next, Section~\ref{sec:understanding} analyzes the dynamics and importance of users in terms of mobility for our case study.

\subsection{Exemplifying the spatio-temporal model}\label{sec:illustrating}
% \subsection{Exemplifying the spatio-temporal model}\label{sec:illustrating}

% Como explicado na seção \ref{sec:DynamicNode2Vec}, nós geramos um espaço vetorial de baixa dimensão preservando a dimensão espaço-temporal através de sequência de redes modeladas usando o \textit{DynamicNode2Vec}. Para ilustrar como nossa proposta funciona, nós usamos às seis primeiras janelas de tempo como exemplo antes de investigar os aspectos de mobilidade no cenário avaliado de forma geral. Relembre ainda, que a parametrização adotada pelo \textit{DynamicNodeVec} produziu uma representação dos nós em um espaço de 50 dimensões. Para observar a disposição dos nós em um espaço vetorial de duas dimensões e exemplificar a sua mobilidade, nós usamos o t-SNE, uma técnica de redução de dimensionalidade para exploração e visualização de dados \cite{Van:2008}. 

To exemplify how our proposal works, we pick just the first six time windows. Recall the \textit{DynamicNodeVec} produce a node embedding representation of 50 dimensions. To be human-visible, we reduce from 50 to 2 dimensional representation by using t-SNE~\cite{Van:2008}, a dimensionality reduction technique for data visualization. Furthermore, to illustrate how the node dynamics can be captured in the latent space modeled, we select only two nodes that presented the highest and lowest average day-to-day mobility (review Section~\ref{sec:DynamicNode2Vec}) according to the first six time windows. We compute the cosine distance day after day for all nodes. Then, we compute the average of these distances for each node. %The nodes with the lowest and highest average mobility in this period had an average cosine distance 0.10 and 0.43 respectively.

\begin{figure}[t]
  \centering
  \includegraphics[height = 5.2cm]{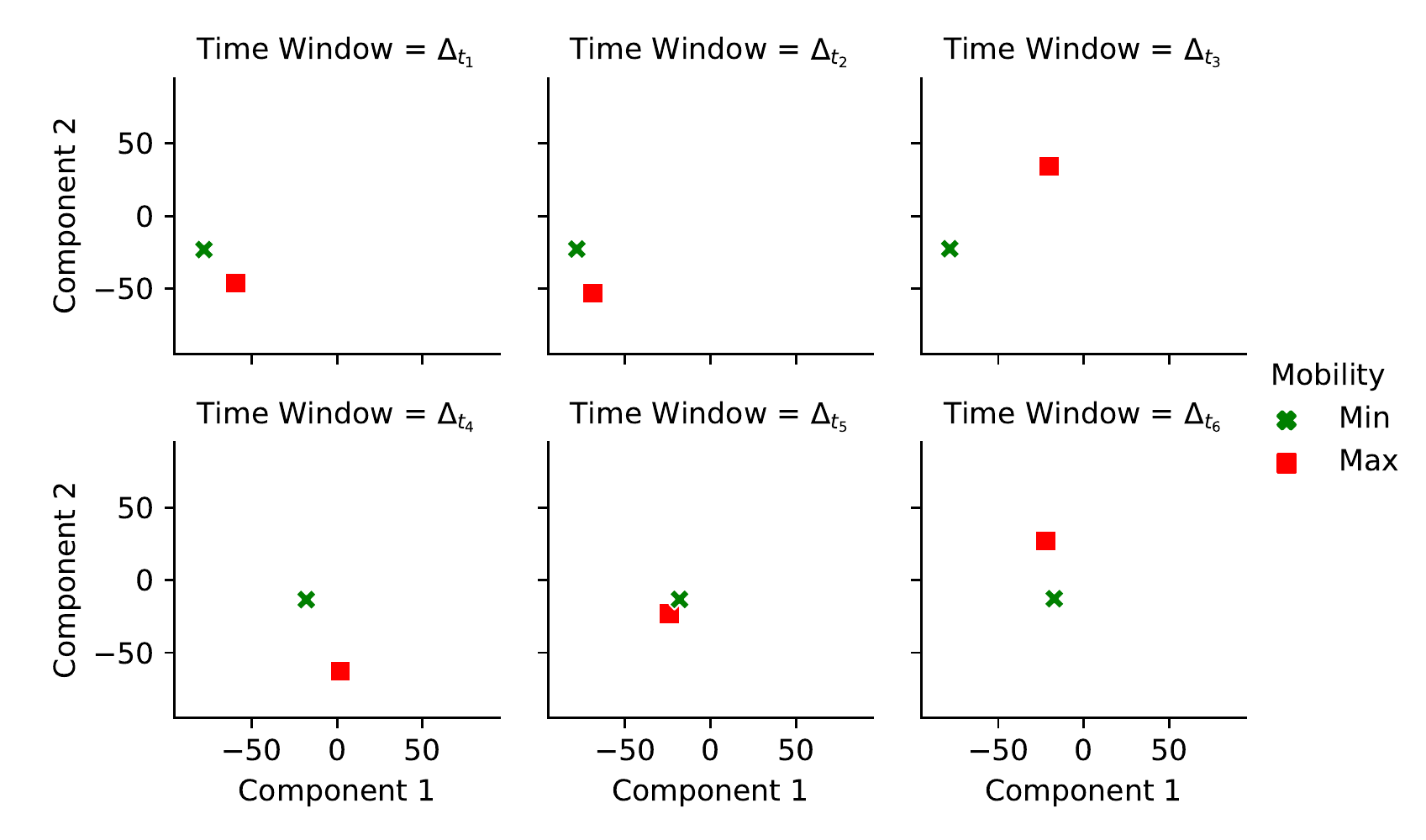} 
  \caption{t-SNE Two-dimensional representation of node embeddings for the first six-time windows.}
%   \caption{2-D representation of the vector space obtained for the first six time windows analyzed.}
   \label{fig:TSNE2} 
\vspace{-0.4cm}
\end{figure}

% A Figura \ref{fig:TSNE2} mostra a representação em 2-D para as seis primeiras janelas de tempo para o nó com maior e menor mobilidade durante este período. Focando no nó que apresentou a menor média (min) podemos observar nas três primeiras janelas de tempo uma certa estabilidade visto que ele se encontra praticamente na mesma posição se observarmos os eixos x e y das três figuras. Entre as janelas de tempo $t_3$ e $t_4$ um deslocamento é observado indicando uma mudança de conectividade deste nó. Por fim, uma estabilidade até a janela $t_6$ é novamente vista. Por outro lado, ao analisarmos o nó com a maior mobilidade média (max) nessas seis primeiras janelas de tempo, é possível notar uma grande variabilidade dia após dia. Dessa forma podemos distinguir os nós com diferentes níveis de mobilidade e, caso necessário, separá-los em períodos específicos conforme a necessidade das aplicações.

Figure~\ref{fig:TSNE2} displays the nodes that have the highest and lowest average mobility in the period. Their average cosine distance were 0.10 (min) and 0.43 (max), respectively. The node with the lowest mobility presented stable (few topological changes) behaviors in the first three time windows (Figure~\ref{fig:TSNE2} top plots). However, between the time windows $\Delta{t_3}$ and $\Delta{t_4}$, a shifting is observed, indicating a change in connectivity of this node. Afterward, the node remains stable until the window $\Delta_{t_6}$. On the other hand, it is possible to notice a high day-to-day variability for the node with the highest average mobility in the period analyzed. In practice, this indicates that this node has significant neighborhood changes in the network. In this way, we can distinguish nodes with different levels of mobility and, if necessary, separate them into specific periods as needed by several network applications.
% Figure \ref{fig:TSNE2} shows the 2-D representation for the first six-time windows for the two nodes. Focusing on the node that presented the lowest average (min), a stable behavior is observed in the first three time windows when observing the x and y axes of the top three figures. However, between the time windows $\Delta{t_3}$ and $\Delta{t_4}$, a shifting is observed, indicating a change in connectivity of this node. Afterward, the node remains stable until the window $\Delta_{t_6}$. On the other hand, it is possible to notice a large day-to-day variability for the node with the highest average mobility (max) in these first six time windows. In practice, this indicates that this node has significantly changed its topology in the network. In this way, we can distinguish nodes with different levels of mobility and, if necessary, separate them into specific periods as needed by several network applications.

% After illustrating the proposed spatio-temporal representation, we deepen our analysis for the whole period studied here, looking at a general understanding of user dynamics in the next section.

\subsection{Unveiling the mobility and connectivity importance}\label{sec:understanding}

In order to understand the mobility patterns, we start by investigating the mobility level of all nodes throughout the period observed on the latent space modeled. Some of the possible applications of our approach include identifying nodes with high mobility degrees for routing in opportunistic or delay-tolerant networks \cite{Sobin:2016, Nunes:2018, Guo:2021}. Therefore, we analyze the nodes' mobility levels assuming that new connections - represented by topological changes in the network and captured in the latent space - can occur with an acceptable delay. Thus, we calculate the cosine distance of a time window $\Delta{t_i}$ for all subsequent time windows $\Delta{t_j}$, such that, $i<j$. In other words, we evaluate the degree of mobility and the potential that a node has to establish new connections throughout the analyzed period.

\begin{figure}[t]
    \begin{subfigure}{0.49\columnwidth}
        \centering
        \includegraphics[width=1\columnwidth]{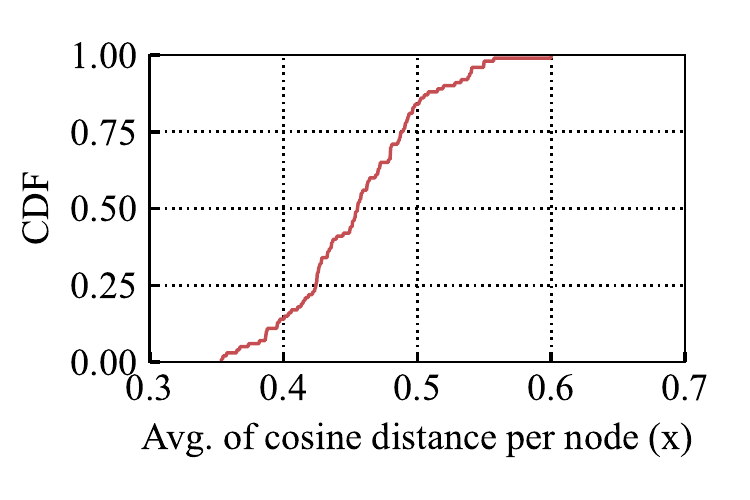}
        \caption{}%Distribuição da média da distância de cosseno
        \label{fig:CDFAVGDC}        
    \end{subfigure}
      \begin{subfigure}{0.49\columnwidth}
        \centering
       \includegraphics[width=1\columnwidth]{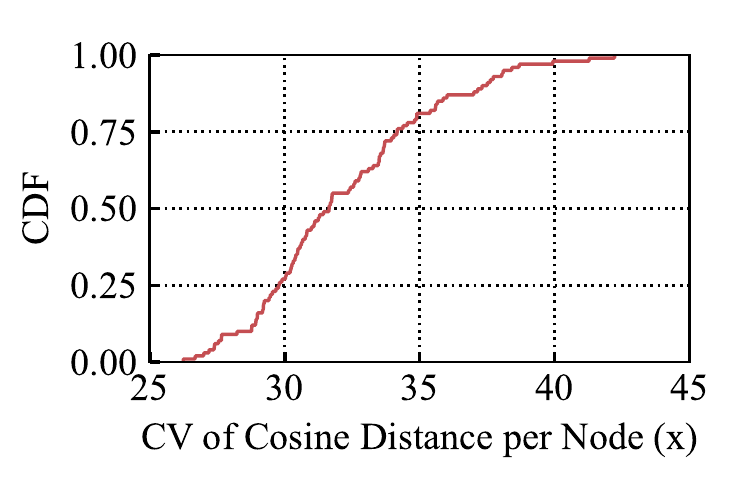}
        \caption{}%Distribuição do CV da distância de cosseno
        \label{fig:CDFCPCVDC}
    \end{subfigure}
    \caption {\gls{cdf} of average and CV of mobility shifting of nodes (measured w.r.t. cosine distance).}
    \label{fig:CDFCVAVGDC}
    \end{figure}

Figure \ref{fig:CDFAVGDC} shows the nodes cosine distance average in a \gls{cdf}. Looking at the 10\% of the most stable nodes, in terms of connectivity, the average cosine distance is at most 0.38. On the other hand, the 10\% of the nodes that change the most shows a mobility average greater than 0.54, exhibiting a 42\% higher mobility level. In order to check to what extent the mobility changes of the nodes are stable or characterized by bursts in specific periods, we compute the \gls{cv} as a measure of dispersion\footnote{\gls{cv} is computed by the ratio of the standard deviation to the mean and expressed in percentage.}. The \gls{cv} distribution is shown in Figure~\ref{fig:CDFCPCVDC}. In practice, values greater than 30 are considered high coefficients of variation~\cite{Reed:2002}. Therefore, it is possible to observe that more than 75\% of the nodes in the network show a greater dynamic in specific periods.
% Figure \ref{fig:CDFAVGDC} shows the cumulative distribution function (CDF) of the cosine distance average of the nodes in the network. Looking at the 10\% of the most stable nodes in terms of connectivity, we have an average cosine distance of at most 0.38. On the other hand, the 10\% of the nodes that change the most show a mobility average greater than 0.54, exhibiting a 42\% higher mobility level. In order to check to what extent the mobility changes of the nodes are stable or characterized by bursts in specific periods, we compute the coefficient of variation (CV) as a measure of dispersion\footnote{CV is computed by the ratio of the standard deviation to the mean and expressed in percentage.}. The CV distribution is shown in Figure \ref{fig:CDFCPCVDC}. Values greater than 30, in practice, are considered high coefficients of variation \cite{Reed:2002}. Therefore, it is possible to observe that more than 75\% of the nodes in the network show a greater dynamic in specific periods.

  \begin{figure}[t]
     \begin{subfigure}{0.495\columnwidth}
        \centering
       \includegraphics[width=1\columnwidth]{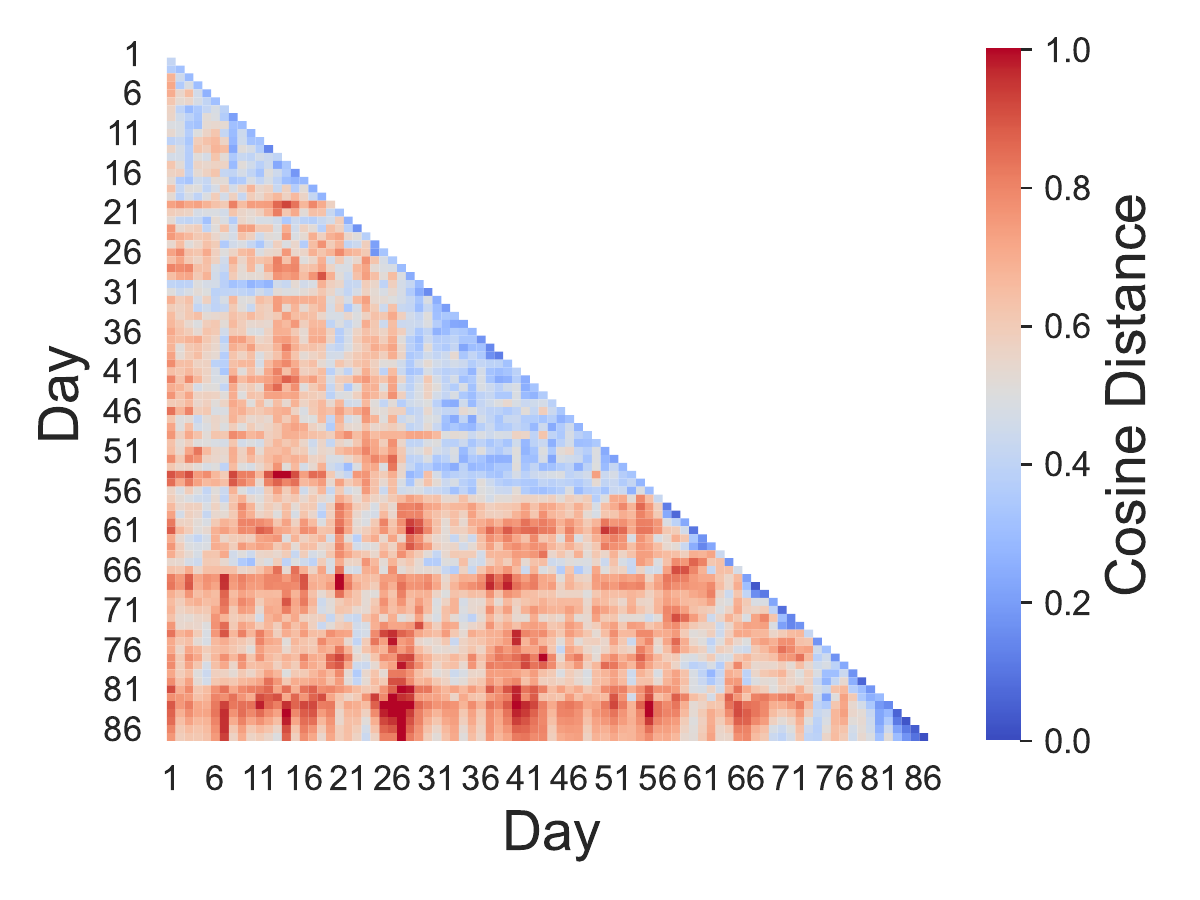}
        \caption{Highest avg. cosine distance}
        \label{fig:HPHMEANDC}
   
    \end{subfigure}
    % \begin{subfigure}{0.49\columnwidth}
    %     \centering
    %   \includegraphics[width=1\columnwidth]{img/P_Heatmap_of_person_with_the_highest_cv_of_the_distance_of_cosine.pdf}
    %     \caption{Nó com maior CV}
    %     \label{fig:HPHCVDC}
    % \end{subfigure}%\\
     \begin{subfigure}{0.495\columnwidth}
        \centering
        \includegraphics[width=1\columnwidth]{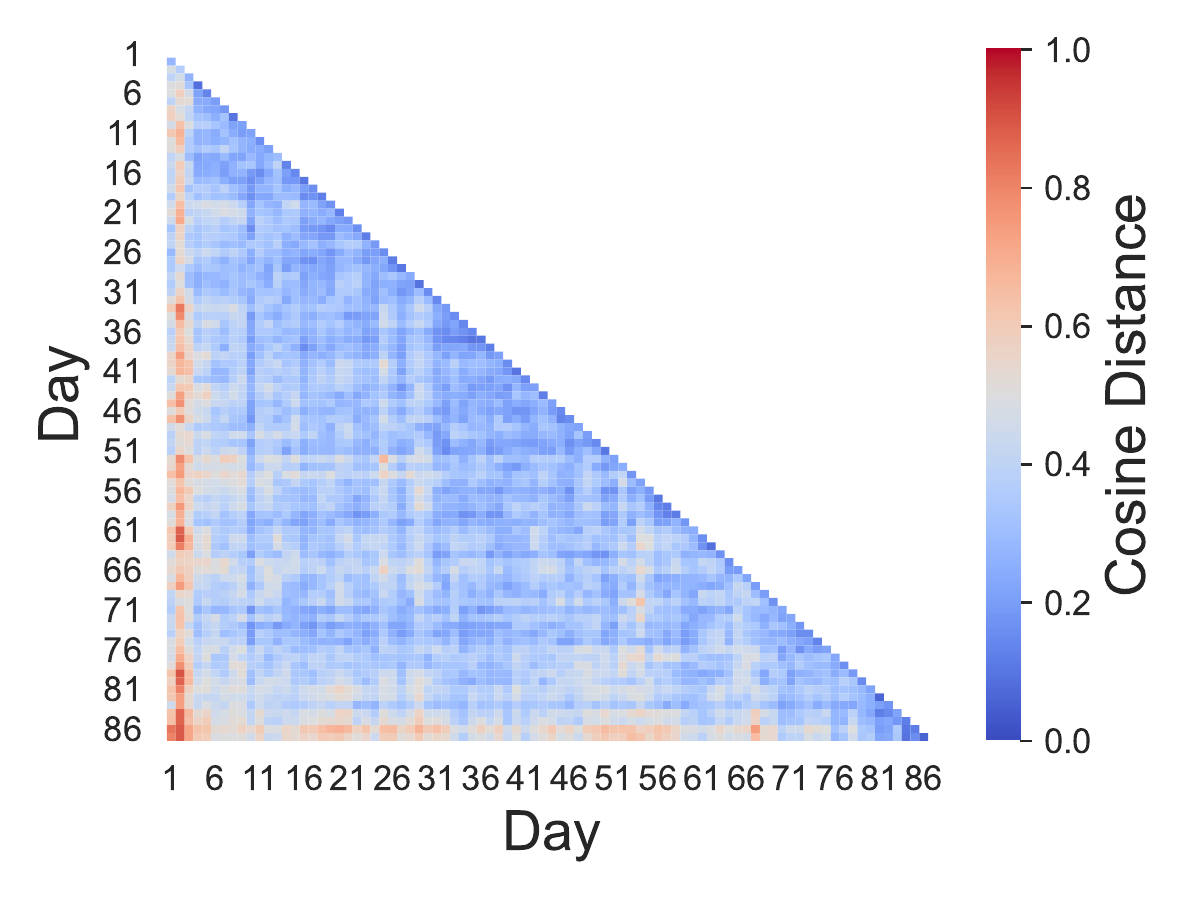}
        \caption{Lowest avg. cosine distance }
        \label{fig:HPLMEANDC}        
    \end{subfigure}
    % \begin{subfigure}{0.49\columnwidth}
    %     \centering
    %     \includegraphics[width=1\columnwidth]{img/P_Heatmap_of_person_with_the_lowest_cv_of_the_distance_of_cosine.pdf}
    %     \caption{Nó com menor CV}
    %     \label{fig:HPLCVDC}        
    % \end{subfigure}
    \caption{Mobility shifting of the nodes with lowest and highest cosine distance over consecutive days.}
    \label{fig:HPMEANDC} 
\end{figure}

We then select nodes with the highest and lowest average cosine distance to investigate mobility changes over consecutive days. Figures \ref{fig:HPHMEANDC} and \ref{fig:HPLMEANDC} show heatmaps where one is able to compare day to day changes over the observed period. The axes $x$ and $y$ indicate the cosine distance to two given time windows. The color indicates the topological change level encoded in the latent space, revealing their degree of mobility according to the cosine distance. The more red and intense the cell, the greater the difference in node connections over the two days observed. Conversely, the more blue and intense, the smaller this difference. It is possible to notice that the node with the high average cosine distance (Figure~\ref{fig:HPHMEANDC}) tends to present high mobility levels for most pairs of days observed, which indicates constant topological changes. We also investigated the node's behavior according to the highest and lowest average CV and found that they exhibit similar behavior at mobility levels to the cosine distance average but with more dispersed changes in even more specific periods.

\begin{figure}[t]
    \begin{subfigure}{0.49\columnwidth}
        \centering
        \includegraphics[width=1\columnwidth]{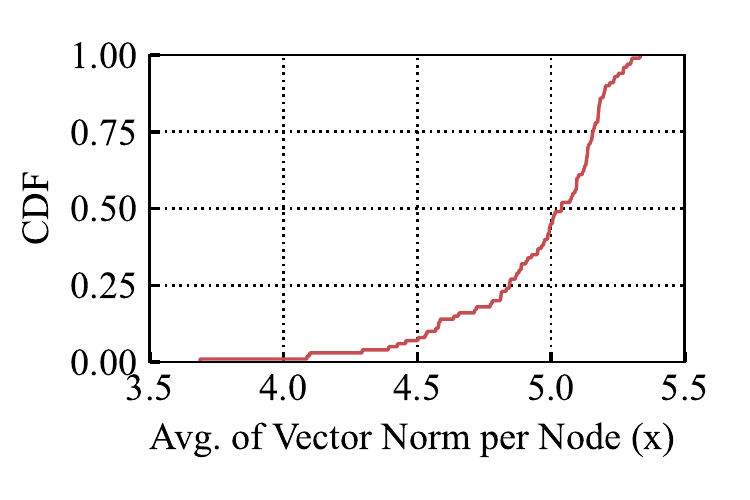}
        \caption{}
        \label{fig:CDFAVGNM}
    \end{subfigure}
    \begin{subfigure}{0.49\columnwidth}
        \centering
        \includegraphics[width=1\columnwidth]{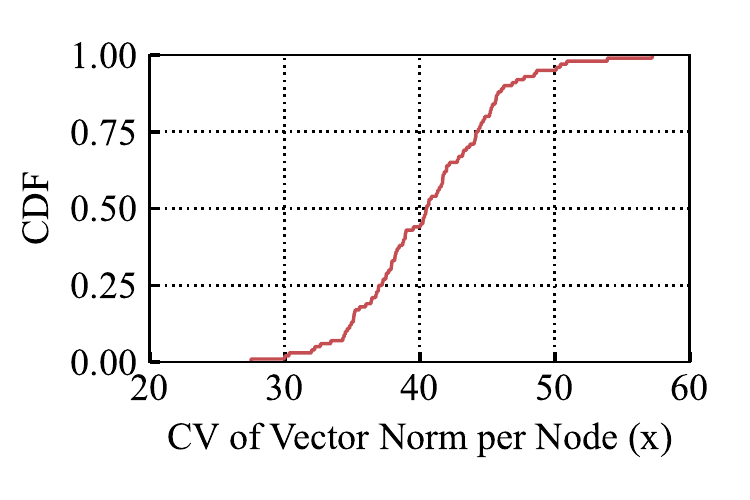}
        \caption{}
        \label{fig:CDFCVNM}
    \end{subfigure}
\caption {\gls{cdf} of average and \gls{cdf} of the nodes' connection importance (measured w.r.t. nodes' vector norm).}

\label{fig:CDFCVAVGNM}
\vspace{-0.4cm}
\end{figure}

% The cosine distance provides an indication of node mobility in terms of connectivity change. However, we cannot say that a node that changes its connections a lot necessarily provides the most important connections for a message to reach the entire network.
% In the context of opportunistic networks, this involves social properties. For example, some people relate to many other people, but these many people can also relate to others, and so on. In this case, such groups can present a very high dynamicity, making it possible to send and receive messages through different connections. Note that the network nodes in this case tend to have a high cosine distance. In another scenario, a group of people may interact a lot with each other and little with outside groups. Also, this little interaction can be stable (always the same people) and be performed by specific members. This makes the other part of this group exclusively need the members with external connections to receive or send messages. In a scenario like this, these \textit{bridges} do not have a high level of mobility, as the external connections are always the same. Therefore, they tend to have a low cosine distance, but they have fundamental connections for a message to reach a certain region. For this reason we evaluate the vector norm obtained for the nodes during the analyzed period, which, as explained in the section \ref{sec:DynamicNode2Vec}, captures the importance of a node in terms of connectivity.

Recall that the cosine distance indicates node mobility in terms of connectivity change. However, we can not claim that a node with a higher mobility level provides the most important connections, for example, to allow a message to reach the entire network. For this reason, we evaluate the vector norm (see details in Section~\ref{sec:DynamicNode2Vec}), which captures the importance of a node in terms of connectivity. Figure~\ref{fig:CDFAVGNM} shows the \gls{cdf} of the average vector norm for all nodes. The 10\% with the lowest value has an average norm smaller than or equal to 4.45. Conversely, the 10\% with the highest average value present an average norm greater than or equal to 5.3, resulting in a difference of almost 20\%. Note that nodes with higher vector norm are more important and central in terms of connectivity. We also evaluate the \gls{cv} of the norms obtained per node and present the distribution in Figure~\ref{fig:CDFCVNM}. Unlike cosine distance, here, 98\% of nodes have a high \gls{cv} (greater than 30). This great variability suggests that some nodes are more important for exchanging information in the network at specific times as network connections evolve.

% Recall that the cosine distance indicates node mobility in terms of connectivity change. However, we cannot claim that a node with a higher mobility level provides the most important connections for allowing a given message to reach the entire network. For this reason, we evaluate the vector norm obtained for the nodes during the analyzed period, which, as explained in the section \ref{sec:DynamicNode2Vec}, captures the importance of a node in terms of connectivity. Figure \ref{fig:CDFAVGNM} shows the distribution of the average for the nodes vector norm. The 10\% with the lowest value has an average norm smaller than or equal to 4.45. Recall that nodes with higher are more important and central in terms of connectivity. Conversely, the 10\% with the highest average value present an average norm greater than or equal to 5.3, resulting in a difference of almost 20\%. We also evaluate the coefficient of variation of the norms obtained per node and present the distribution in Figure \ref{fig:CDFCVNM}. Unlike cosine, 98\% of nodes have a high CV (greater than 30). This great variability suggests that some nodes are more important for exchanging information in the network at specific times.      

\begin{figure}[t]
\centering
\includegraphics[height=4.5cm]{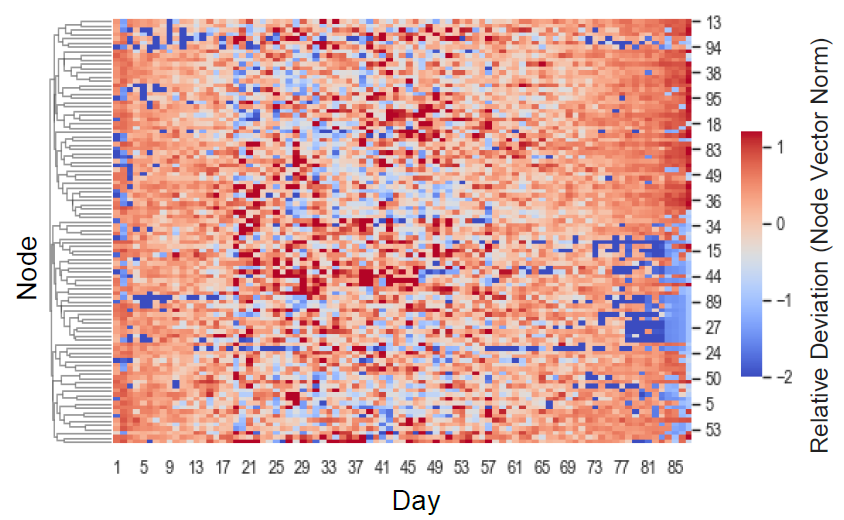} 
\caption{Heatmap of the node vector norm over the observed period. }
\label{fig:CLUSTERMAPNORMA} 
\vspace{-0.4cm}
\end{figure}

The vector norm of each node in each day can be used to understand nodes' connectivity importance in the temporal aspect. Figure~\ref{fig:CLUSTERMAPNORMA} presents a day-to-day heatmap for all nodes' norm vectors. Rows represent the nodes and columns represent the days. Each day was normalized by using the \textit{z-score}\footnote{Each value is subtracted by the column average and divided by the column standard deviation.}. Thus, one can see how much node's importance deviates relatively from the average importance over days. The color of each cell reflects the normalized norm value. Also, we compute a dendrogram to group nodes that present similar importance on specific days. The Pearson correlation between the node norms was employed as dendrogram grouping metric~\cite{Benesty:2009}. We can observe that some nodes present greater importance than the others (indicated by red cells) in specific periods, confirming our finding about the high \gls{cv} of the norm. %Furthermore, the dendrogram reveals subsets of nodes with similar importance on specific days.
% In order to understand the nodes' importance connectivity considering the temporal dimension, we present a heatmap with the norm of each node for each day of the  period analyzed in Figure \ref{fig:CLUSTERMAPNORMA}. Rows represent the nodes and columns represent the days. Each column (day) was normalized using the \textit{z-score}\footnote{Each value is subtracted by the column average and divided by the column standard deviation.}. This way, we can see how much node importance deviates relatively from the average importance of all nodes in the network for each day. The color of each cell reflects the normalized norm value. Finally, to measure the similarity of nodes in terms of importance, we compute a dendrogram under the rows, aiming at grouping nodes that have equal importance on specific days. The Pearson correlation between the node norms was used as the dendrogram metric \cite{Benesty:2009}. In general, we can observe that some nodes present greater importance than the others (indicated by red cells) in specific periods, confirming our finding about the high CV of the norm. Furthermore, the computed dendrogram reveals subsets of nodes with similar importance on specific days.

As the cosine distance and the vector norm provide complementary analyses, we investigate how these measures are related. Intuitively, we aim to understand how the mobility of nodes is related to the importance of their connections. To do so, we apply the Pearson's correlation between the average and \gls{cv} of both measures, the result is summarized in Table~\ref{tab:correlation}. The strongest correlation is negative and occurs between the average of the norm and the \gls{cv} of the cosine distance. This suggests that nodes with more important connections have a stable mobility pattern. Then, we observe a moderate correlation between the average cosine distance and the \gls{cv} of the norm, which means that nodes who consistently make abrupt connectivity changes tend to maintain connections that are sometimes important and sometimes not. The other correlations are not significant for $p=0.01$.

\begin{table}[t]
\centering
\tiny
\caption{Correlations between cosine distance and the vector norm.}
\label{tab:correlation}
\begin{tabular}{|c|c|c|}
\hline
\multicolumn{2}{|c|}{\textbf{Metrics}}                        & \textbf{Correlation} \\ \hline
Avg. of Cosine Distance & CV  of Cosine Distance & -0.04      \\ \hline
Avg. of Cosine Distance & CV of Vector Norm             & 0.70       \\ \hline
Avg. of Vector Norm            & CV of Cosine Distance & -0.95      \\ \hline
Avg. of Vector Norm            & CV of Vector Norm             & -0.10      \\ \hline
\end{tabular}
\end{table}

We also investigate the Person's correlation between our approach and traditional graph topological importance measures~\cite{Daly:2007, Wan:2017, Zhang:2017, Nunes:2018, Guo:2021}. To do it, we consider the main well-known topological measures used in these works. Remember, they capture the most varied notions of nodes' centrality and connectivity in the network, and, therefore, different notions of importance. Table~\ref{tab:Correlacoes2}\footnote{All observations are significant for $p=0.01$.} depicts the correlations between these measures.
% Finally, we close our analysis by exterminating whether topological measures of the graph are typically used in mobile network strategies \cite{Daly:2007, Wan:2017, Zhang:2017, Nunes:2018, Guo:2021} can capture the patterns of connectivity extracted by our methodology and vice-versa. To do it, we consider the main topological measures used in these works. Remember, they capture the most varied notions of centrality and connectivity of nodes in the network and, therefore, different notions of importance. Table \ref{tab:Correlacoes2}\footnote{All observations are significant for $p=0.01$.} depicts the correlations between these measures.

Turning the focus on the centrality measures (Degree, Betweenness, Closeness, and Eigenvector). It is possible to observe a moderate negative correlation between traditional graph measures and the average of the cosine distance. It suggests that the more dynamic a node is, the lower its centrality according to different views. The average vector norm presents slightly positive correlation with respect to traditional measures, a possible explanation is that  centrality measures typically capture unique notions of centrality and importance, and do not consider the temporal factor~\cite{Valente:2008}. Hence, they cannot generalize a concept of centrality that is universally important in the graph. Finally, we consider the clustering coefficient, which shows a low-to-moderate positive and negative correlation, respectively, with both node metrics in latent space. The positive case suggests that nodes with neighbors more connected (high clustering coefficient) tend to change their connections collectively (high avg. of cosine distance), which may reflect our case study's regular group meetings property. On the other hand, nodes with less-connected neighborhoods tend to have essential links for network connectivity (high avg. of vector norm), explained by bridges and peripheral connections in the networks. %In short, the mobility patterns revealed by our approach moderately aggregate different notions of importance about the connectivity of a node in the network simultaneously.

\begin{table}[t]
\centering
\tiny
\caption{Correlations between topological and embedding measurements.}
 \label{tab:Correlacoes2} 
\begin{tabular}{|c|c|c|}
\hline
\textbf{Embedding} & \textbf{Topological}   & \textbf{Correlation} \\ \hline
Avg. of Cosine Distance         & Avg. of Degree                   & -0.64               \\ \hline
Avg. of Cosine Distance         & Avg. of Betweenness & -0.69               \\ \hline
Avg. of Cosine Distance         & Avg. of Closeness            & -0.64               \\ \hline
Avg. of Cosine Distance         & Avg. of Eigenvector              & -0.65               \\ \hline
Avg. of Cosine Distance        & Avg. of Clustering Coefficient   & 0.51                \\ \hline
Avg. of Vector Norm                 & Avg. of Degree                   & 0.33                \\ \hline
Avg. of Vector Norm                 & Avg. of Betweenness & 0.50                \\ \hline
Avg. of Vector Norm                 & Avg. of Closeness           & 0.28                \\ \hline
Avg. of Vector Norm                 & Avg. of Eigenvector              & 0.32                \\ \hline
Avg. of Vector Norm                 & Avg. of Clustering Coefficient   & -0.55               \\ \hline
\end{tabular}
\end{table}

\section{Conclusion}\label{sec:conclusion}

Understanding mobility in mobile networks includes examining how entities move along with their role for network connectivity. Those are patterns notably relevant to design networking protocols and solutions. This work proposed a methodology for modeling and analyzing mobile networks by employing a node embedding approach capable of unveiling patterns related to network connectivity, mobility, and their natural evolution. We noticed that most solutions designs in the literature rely on graph theory metrics computer on the network. Consequently, they tend to privilege particular characteristics indicating which nodes are central points in the structure. Alternatively, our approach offers a generalized notion of connectivity importance based on the existing links in each node's network tracking it while the connections in the network change.
% Understanding mobility in mobile networks includes examining how entities move along with their role for network connectivity. Those are patterns notably relevant to design networking protocols and solutions. This work proposed a methodology for modeling and analyzing mobile networks by employing a node embedding approach capable of unveiling patterns related to network connectivity, mobility, and their natural evolution. We noticed that most designed solutions in the literature rely on graph theory metrics computer on the network. Consequently, they tend to privilege particular characteristics indicating which nodes are central points in the structure. Alternatively, our approach offers a generalized notion of connectivity importance based on the existing links in each node's network tracking it while the connections in the network change.

By applying our methodology in a case study as a proof of concept, we showed that our approach could reveal nodes with different mobility and connectivity importance levels, whereas the network topology evolves. Finally, we contrasted the notion of connectivity importance provided by our methodology with some topological network measurements used network solutions present in the literature. We then observed that these measures could not generalize the patterns of connectivity captured by our methodology.

As future works, our methodology offers a range of possibilities. An immediate opportunity is to apply our methodology to design new solutions that require an efficient node selection strategy for dissemination/collection of information, location, vehicular communication, among other applications in the context of smart cities. It also includes evaluating the performance of such protocols, comparing purely topological measures to those proposed here. Finally, another possibility is to extend the network modeling to include contextual information about the nodes' visitation locations. It would enable new tasks to be performed, including prediction and recommendation of points of interest to users.

\bibliographystyle{abbrv}

 \bibliography{references}

\end{document}